\documentclass[10pt]{article}
\usepackage{graphicx}
\usepackage{dcolumn}

\begin{document}

\title{Who Wants to Self-Track Anyway? Measuring the Relation between Self-Tracking Behavior and Personality Traits}

\author{
        Georgios~Chatzigeorgakidis \\
        Department of Informatics and Telecommunications\\
        University of Peloponnese, Tripolis, Greece\\
        chgeorgakidis@uop.gr
        \and
        Andrea Cuttone\\
        DTU Compute\\
        Technical University of Denmark, Kgs. Lyngby, Denmark\\
        ancu@dtu.dk
        \and
        Sune Lehmann\\
        DTU Compute\\
        Technical University of Denmark, Kgs. Lyngby, Denmark\\
        sljo@dtu.dk
        \and
        Jakob Eg Larsen\\
        DTU Compute\\
        Technical University of Denmark, Kgs. Lyngby, Denmark\\
        jaeg@dtu.dk
}
\date{\today}

\maketitle

\begin{abstract}
We describe an empirical study of the usage of a mobility self-tracking app, \textit{SensibleJournal 2014}, which provides personal mobility information to $N$=796 participants as part of a large mobile sensing study. Specifically, we report on the app design, as well as deployment, uptake and usage of the app. The latter analysis is based on logging of user interactions as well as answers gathered from a questionnaire provided to the participants. During the study enrollment process, participants were asked to fill out a questionnaire including a Big Five inventory and Narcissism NAR-Q personality tests. A comparison of personality traits was conducted to understand potential differences among the users and non-users of the app. We found a relation between self-tracking and conscientiousness, but contrary to the view in popular media, we found no relation between self-tracking behavior and narcissism.
\end{abstract}

\section*{Introduction}
\label{sec:introduction}
Recently, the area of lifelogging, Quantified Self and Personal Informatics have gained substantial attention and uptake due to the availability of smartphones, low-cost wearable sensors and, more recently, smart watches. 
This development has significantly lowered the barrier for people to engage in a wide range of self-tracking activities, with monitoring of exercise, physical activity, and step counting being widely adopted. 
Also in research, the area has gained increased attention in recent years, with international workshops on Personal Informatics \cite{froehlich2014disasters,li2013personal, li2012personal}, as well as conferences having sessions on Quantified Self. 
However, paradoxically, empirical research describing the self-tracking phenomenon is somewhat limited \cite{ruckenstein2014visualized, bode2015}.

In this paper we describe an empirical study of mobility self-tracking, using a smartphone app that has been developed as part of our research. 
We measured the usage of the self-tracking app on a population of almost 800 individuals -- bachelor level university students from all technical sciences, for a duration of four months. 
In an ongoing mobile sensing study~\cite{sensibleDTU_paper} we offered all participants a self-tracking app that provides a feedback interface reporting on personal mobility patterns on a daily basis. 
All participants were informed about the existence of the app through a notification system. 
While all participants were instructed to install the app on their smartphone, usage of the app was optional. 

The motivation behind the study was to measure the uptake of a Quantified Self app in order to see how many would be interested in self-tracking, as well as the duration of the interest in the app, in terms of usage patterns over time. 
As the participants were enrolled in the mobile sensing study at the time of the introduction of the app, mobility and social interaction data were already being collected. 
Thus, the participants were merely provided with a mobile feedback interface providing access to personal data that was already being collected about them.  

As it has been debated whether specific personality traits are distinctive among self-trackers, we wanted to measure the relation between self-tracking behavior and personality traits. 
In particular, the \textit{narcissism} label has  been associated with self-tracking in popular media and has been debated in research literature too~\cite{bode2015, lupton2014selfcultures}.
% We hypothesize that narcissism personality traits and self-tracking behavior are uncorrelated.

\section*{Related work}
Li et al.~\cite{stage_based} describe Personal Informatics using a model that involves a five stage process, where a key stage is \textit{self-reflection}, which is often supported through visualization of the collected personal data. 
Data visualization is seen as a mean to gain insights into personal behaviors, which can shape the basis for achieving behavior change.
%As denoted by \cite{visualization_heuristics},  this role can be played by proper visualizations, designed and implemented according to four suggested heuristics. 

There is a plethora of examples, both scientific and commercial, that attempt to leverage the potential of data visualization as the means for individuals to interact with and gain insights from personal data. 
The most relevant to this study is the 2013 version of SensibleJournal, providing visualizations of mobility patterns \cite{spiral}, as well as visualizations on social interactions of its users \cite{sensiblejournal2013}. 
Other examples that involve visualizing information on a map are \textit{Personal Driving Diary}~\cite{driving_diary}, which presents images of the detected events during driving along with pinpointed locations, and \textit{Now Let Me See Where I Was}~\cite{now_let_me_see}, which creates mobility-based visualizations for each participant by pinpointing locations on a map.

The self-tracking phenomenon has attracted attention in popular media, with articles sometimes suggesting a tendency of self-trackers towards narcissism \cite{wolf2010data}. 
Similarly, in recent report from Symantec on Quantified Self data security, Quantified Self is described as part of ``\textit{a trend towards [...] narcissism}'' \cite{barcena2014}.  

Research literature has discussed this popular media view on the practice of self-tracking as obsessive or narcissistic \cite{lupton2014self, lupton2014selfcultures, lupton2014you} and in \cite{waltz2012quantified}, the initial encounter with self-tracking is being described as something that would appear to be ``\textit{just another example of technology stretching the limits of narcissism}''. 
However, this viewpoint on self-tracking has been criticized by Bode and Kristensen \cite{bode2015}, with a call for a more varied description of the self-tracking phenomenon. 

An attempt to measure narcissism among self-trackers has been made on a small scale (N=36) in an online survey in the Quantified Self community, using the NPI-16 test \cite{ames2006npi} with 16 questions related to narcissism \cite{wolf2009}.
The result was a 0.38 score on narcissism compared to the mean scores in five American studies, which reported to be in the range 0.31-0.41 \cite{ames2006npi}. 
While concluding that there was no correlation between self-tracking and narcissism, it is also suggested that the definition of narcissism may not be clear \cite{wolf2009}.

\section*{Method}
This work is part of the \textit{SensibleDTU} project, a large-scale study of high-resolution social networks, which is described in detail in \cite{sensibleDTU_paper}. 
Data collection was approved by the Danish Data Protection Agency, and informed consent has been obtained for all study all participants.
We study $N$=796 first year students provided with an Android smartphone.
The phone is equipped with a data collector app running continuously in the background. 
The latter collects and periodically uploads data to a server from multiple sources: location, Bluetooth, calls, SMS, and WiFi. 
The participants were also asked to fill out a questionnaire including the Big Five inventory \cite{john1999big} and Narcissism NAR-Q \cite{back2013narcissistic}, from which we can deduce the following six personality traits: \textit{extraversion}, \textit{agreeableness}, \textit{conscientiousness}, \textit{neuroticism}, \textit{openness} and \textit{narcissism}. 
We have found that our population under study is unbiased with respect to the personality traits of the general population \cite{sensibleDTU_paper}.

All participants that had joined the study were requested to install SensibleJournal 2014, a self-tracking mobile app designed to support self-reflection \cite{stage_based} on personal mobility, in terms of places visited and movements between them. 
The locations are extracted by clustering groups of consecutive locations within a predetermined distance, as described in \cite{cuttone2014inferring}. 

\subsection*{The SensibleJournal 2014 App}
\begin{figure}[!t]
\centering
\includegraphics[width=1\columnwidth]{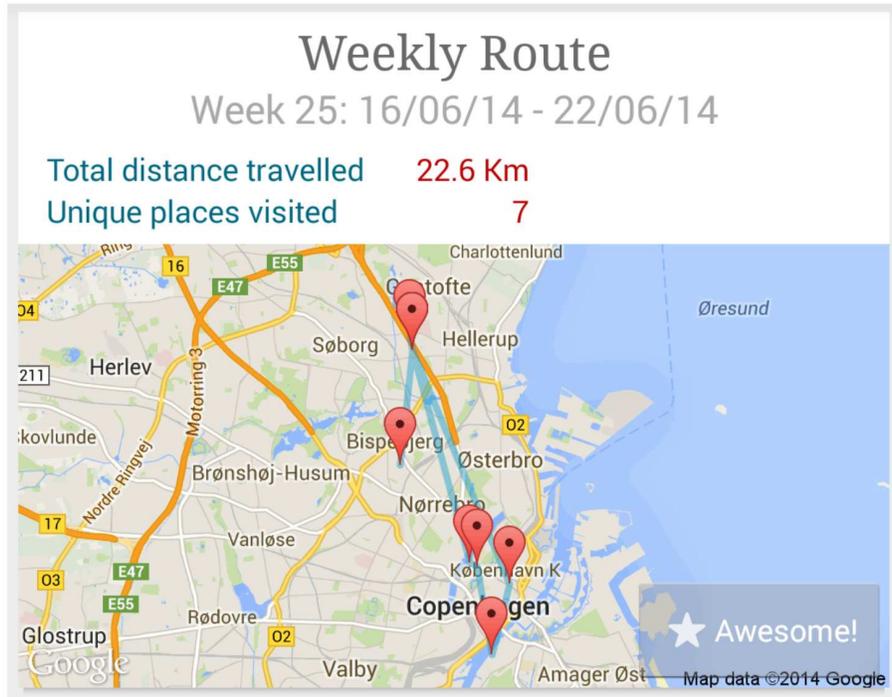}
\caption{An example card showing mobility information in the SensibleJournal 2014 mobile app. 
The app user interface contains a list of such cards showing different views of personal mobility}
\label{main_view}
\end{figure}

The mobile app uses a card-based user interface, which shows mobility related information on cards that appear on a continuous timeline from most to least recent. 
Each card contains a static mini-map, which pinpoints locations of interest, along with specific informative text. 
We provide six different types of cards, each presenting different information about personal mobility: ``\textit{My Current Location}'', ``\textit{Last Visited Place}'', ``\textit{Latest Journey}'', ``\textit{Daily Route}'', ``\textit{Weekly Route}'', and ``\textit{Most Visited Places}''. 
The cards contain a static map along with descriptive text. 
An example card is shown in Fig.~\ref{main_view}.

% NOTE: Added a card figure instead of the whole app screenshot to save space

A card can be tapped in order to open the corresponding detailed view, which offers a more informative visualization through an interactive map which offers the ability to pan and zoom. 
Users can also access their history through an archive view that chronologically lists specific older detailed views, accessible from a ``navigation drawer'' (by tapping on the app title area). 
In order for the participants to be reminded to check for newly available cards, the app sends a periodic reminder using a notification on users’ devices. 
To avoid intrusive behavior, the notification is sent once every three days at noon.

\subsection*{Usage Data Collection}
SensibleJournal 2014 gathers data about the way users interact with the app. 
In particular, the app logs interaction events, such as when a user launches/pauses the app, or navigates through the interface. 
Each card contains an ``\textit{Awesome!}'' button (see Fig.~\ref{main_view}), which can be tapped in order to provide feedback about the cards. 
The usage log is periodically uploaded and stored on our server.

\subsection*{Results}
\begin{figure}[!t]
\centering
\includegraphics[width=1\columnwidth]{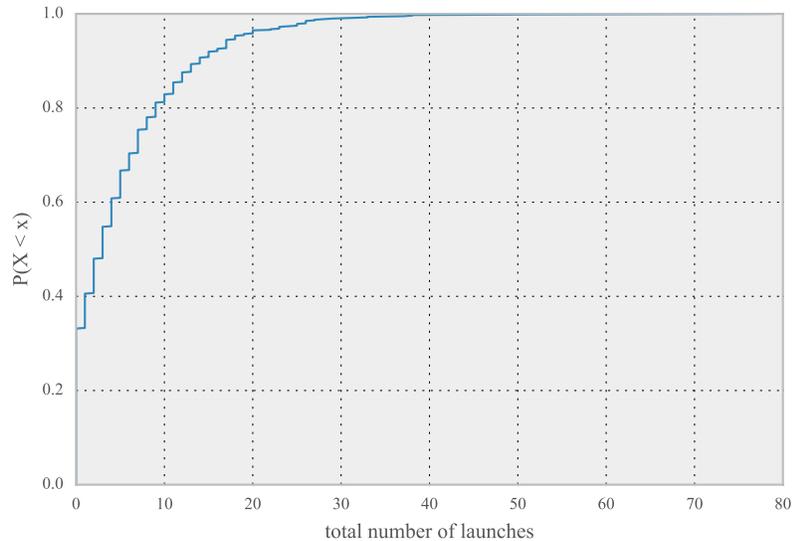}
\caption{Cumulative distribution function of usage (number of app launches). 
The usage decays exponentially, with only a small percentage of participants using the app on a regular basis }
\label{ccdf-time}
\end{figure}

%\subsection*{General Usage}
We analyze the usage logs between late June and October 2014. 
For each participant we consider events that lasted at least 5 seconds and at most 10 minutes to avoid considering accidental app launches, or anomalies in the usage collection. 
In total, 242 (30.4\%) of all 796 participants had no interaction with the app at all. 
Even though they had to install the app when joining the study, they never launched it. 
The cumulative distribution function of the total number of times that the app was launched per user, illustrated in Fig.~\ref{ccdf-time}, shows that the usage decays exponentially: around 60\% of the participants launched the app less than 5 times or not at all and less than 5\% launched it more than 20 times. 
This is in line with our findings in previous work \cite{cuttone2014long}.

\begin{figure}[!t]
\centering
\includegraphics[width=1\columnwidth]{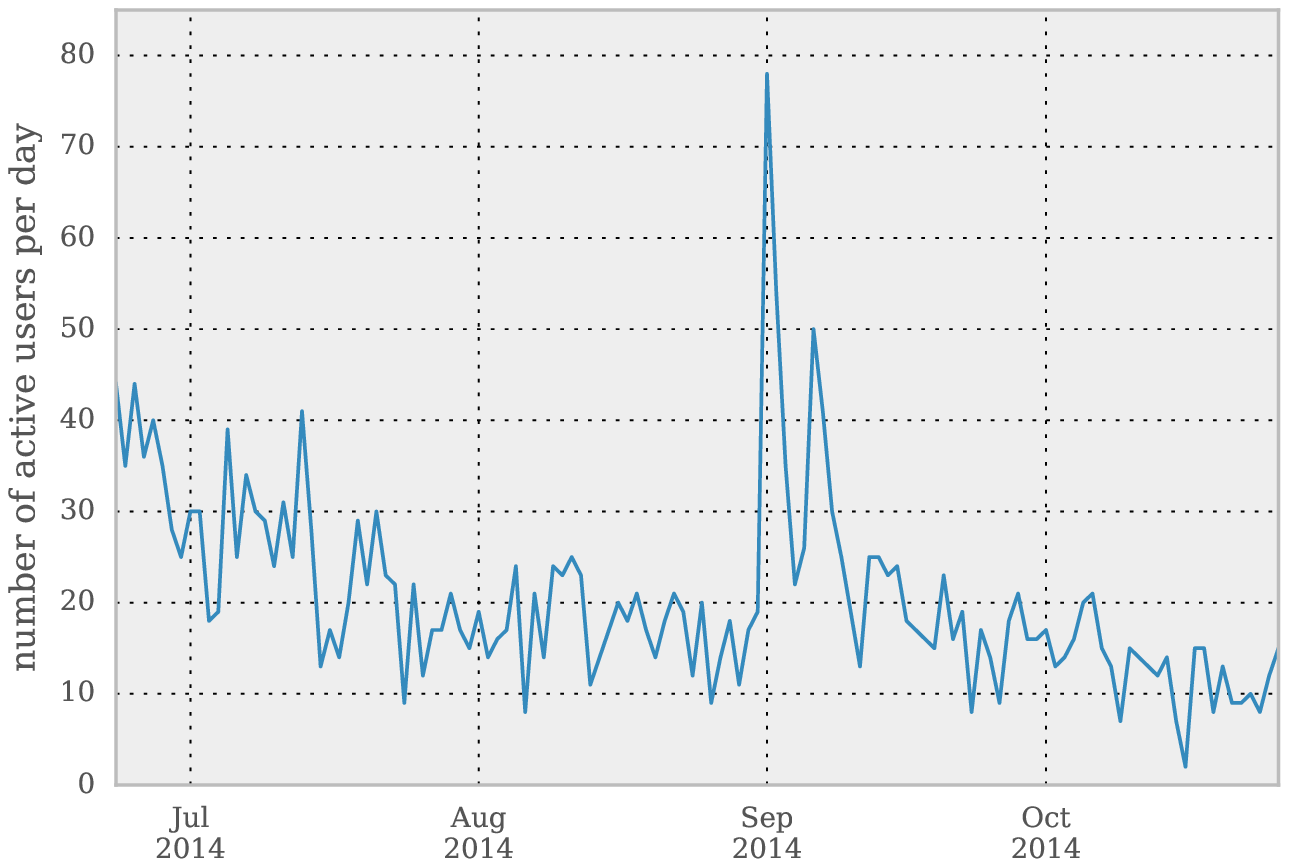}
\caption{The number of active users as a function of time illustrates the decreasing trend from the beginning of the experiment, with an exception being early September, which coincides with the beginning of a new semester}
\label{daily-usage}
\end{figure}

Additionally, we count the per-day number of users with at least one launch (Fig.~\ref{daily-usage}). 
The number of active users slowly decays from the start of the experiment. 
There is a peak in the beginning of September, which coincides with the start of the new university semester. 
A similar decay over time was also reported in previous work \cite{cuttone2014long}.

From the 796 participants, only 16 individuals (2\%) used SensibleJournal 2014 more systematically. 
We defined as more regular users, the ones that used the app at least 20 times and at least once per month during the experiment.

\subsection*{Usage and Notifications}
\begin{figure}[!t]
\centering
\includegraphics[width=1\columnwidth]{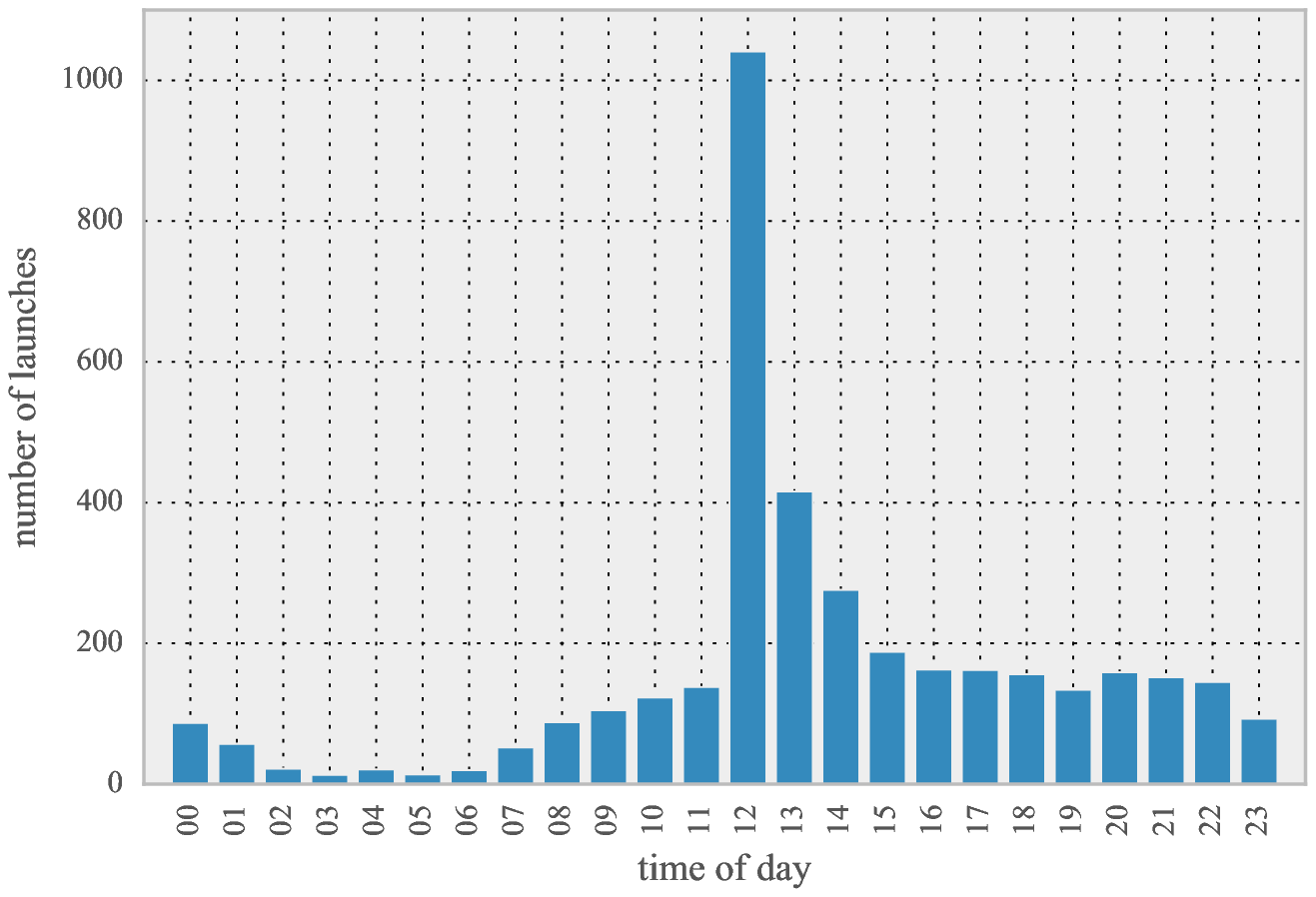}
\caption{Usage per time of day. 
A clear peak happens between 12 and 14, probably due to the notifications scheduled at noon}
\label{hourly-usage}
\end{figure}

As mentioned, a notification system alerted the participants about new cards every three days, at noon. 
Fig.~\ref{hourly-usage} shows that the total number of the app launches was significantly higher between 12:00 and 14:00, which suggests that the notification was an important factor in engaging users (Fig.~\ref{daily-usage}). 
This observation is in line with \cite{bentley2013power}, which reported a notably higher usage of a self-tracking mobile app with reminding through notifications.

\subsection*{Survey}

In December 2014, all the participants were contacted and asked to fill out an electronic survey with the following questions:
\begin{enumerate}
\item Have you discovered something new or interesting about yourself? If yes, what? (open answer)
\item If you no longer use SensibleJournal 2014 app, why not? (multiple options)
\begin{enumerate}
\item The app is too slow
\item The visualizations are confusing
\item I am not interested in my location data
\item I do not know the app
\item The app always shows similar information
\item I do not learn anything new from my data
\end{enumerate}
\item What do you think of the notifications?
\begin{enumerate}
\item There should be more
\item There should be less
\item They should be removed
\end{enumerate}
\item Have you clicked ``Awesome!''? If yes, why?
\begin{enumerate}
\item I liked the visualizations
\item I liked the information shown in the visualizations
\end{enumerate}
\item How do you use the app?  (open answer)
\item Do you have any other comments or suggestions? (open answer)
\end{enumerate}

A total of 51 participants answered the survey (6\% response rate). 
The majority reported not learning anything interesting, however some reported being surprised to gain new knowledge about their daily patterns: ``\textit{I am very surprised about my monotonous patterns, home-work-home}'', ``\textit{Although I thought that I move around a lot, I basically spend my time with a specific friend of mine}'', ``\textit{there is a whole routine in our lives, that is surprising!}'' and ``\textit{I go out at the same places (more or less)}''. 
One participant reported checking the app after a change in his/her routine, and several reported using the app for learning about the time spent between home and university. 
The repetitiveness of daily routine and consequently of the app feedback, was one of the prominent causes for many users to stop using it: 43\% reported not learning anything new and 22\% reported that the app shows similar information every time.
One participant even suggests that the app should provide feedback only when ``\textit{something new happens}'' such as visiting a place never seen before. 
Moreover, 16\% complained that the visualizations were confusing and 28\% said not to be interested in their location data.
Regarding the ``\textit{Awesome!}'' button, 18\% reported to have clicked it because they liked the visualizations and 20\% because they found the information shown in the visualizations useful. 16\% reported that there should be more notifications, 30\% preferred less and 24\% would prefer not to have notifications at all.

\subsection*{Personality}
For each participant we compute the following six personality traits based on the answers to the questionnaire including the Big Five inventory \cite{john1999big} and Narcissism NAR-Q \cite{back2013narcissistic}: extraversion, agreeableness, conscientiousness, neuroticism, openness and narcissism. 
Additionally, for each participant we determine a number of features based on the collected usage data:

\begin{itemize}
\setlength\itemsep{0em}
\item number of days with at least one launch
\item total time interacting with the app
\item total number of launches
\item mean session duration
\end{itemize}

For the users who have no usage data we assigned the value 0 to all the above mentioned features.
We split the population into the top 10\% and remaining 90\% quantiles according to each usage feature and compared the distribution of each personality trait between the top and remaining quantiles using t-tests (6 personality traits x 4 usage features = 24 tests).
Table~\ref{p_values} contains the corresponding $p$-values for each trait and feature pair.

\begin{table}[h!]
\centering
\caption{$p$-values for the t-tests for each trait-feature pair}
\label{p_values}
\begin{tabular}{|m{1.45cm}|m{1.5cm}|m{1.5cm}|m{1.5cm}|m{1.5cm}|m{1.5cm}|m{1.5cm}|}
\hline                      & extra-version & agreeable-ness & conscien-tiousness & neuroti-cism & openness & narcissism \\ \hline
total events          & 0.6970       & 0.5653        & 0.0342            & 0.5141      & 0.2301   & 0.0677    \\ \hline
total time            & 0.6943       & 0.7673        & 0.0009            & 0.5214      & 0.4531   & 0.3882    \\ \hline
mean session dur. 	  & 0.3382       & 0.3485        & 0.4042            & 0.4401      & 0.0614   & 0.7776    \\ \hline
active days           & 0.0416       & 0.3702        & 0.3418            & 0.2676      & 0.0319   & 0.1147 \\ \hline
\end{tabular}
\end{table}

However, after correcting for multiple comparisons using the Holm-Bonferroni method~\cite{holm1979simple}, we find that the only statistically significant difference is between conscientiousness and total time. 
Table~\ref{p_values_corrected} displays the corrected $p$-values.

\begin{table}[h!]
\centering
\caption{$p$-values for each trait-feature pair, after the Holm-Bonferroni correction}
\label{p_values_corrected}
\begin{tabular}{|m{1.45cm}|m{1.5cm}|m{1.5cm}|m{1.5cm}|m{1.5cm}|m{1.5cm}|m{1.5cm}|}
\hline & extra-version & agreeable-ness & conscien-tiousness & neuroti-cism & openness & narcissism \\ \hline
total events          & 1.0000       & 1.0000         & 0.7515            & 1.0000      & 1.0000   & 1.0000    \\ \hline
total time            & 1.0000        & 1.0000        & \textbf{0.0210}            & 1.0000      & 1.0000   & 1.0000    \\ \hline
mean session dur. 	  & 1.0000       & 1.0000        & 1.0000            & 1.0000      & 1.0000   & 1.0000    \\ \hline
active days           & 0.8726        & 1.0000        & 1.0000            & 1.0000      & 0.7338   & 1.0000 \\ \hline   
\end{tabular}
\end{table}

The Holm-Bonferroni correction is quite strict, therefore we provide another view of the results using bootstrap to illustrate the differences.
We calculate the bootstrapping distributions of the means of the top 10\% quantiles, for each personality trait-usage feature pair. 
In particular:
\begin{itemize}
\item We calculate the mean of the trait for the top 10\% according to the usage feature
\item We bootstrap $n = 5000$ subsamples from the initial population and obtain their means
\item We compare the distribution of the means of the subsamples and the mean of the top 10\%
\end{itemize}

\begin{figure}[!t]
\centering
\includegraphics[width=1\columnwidth]{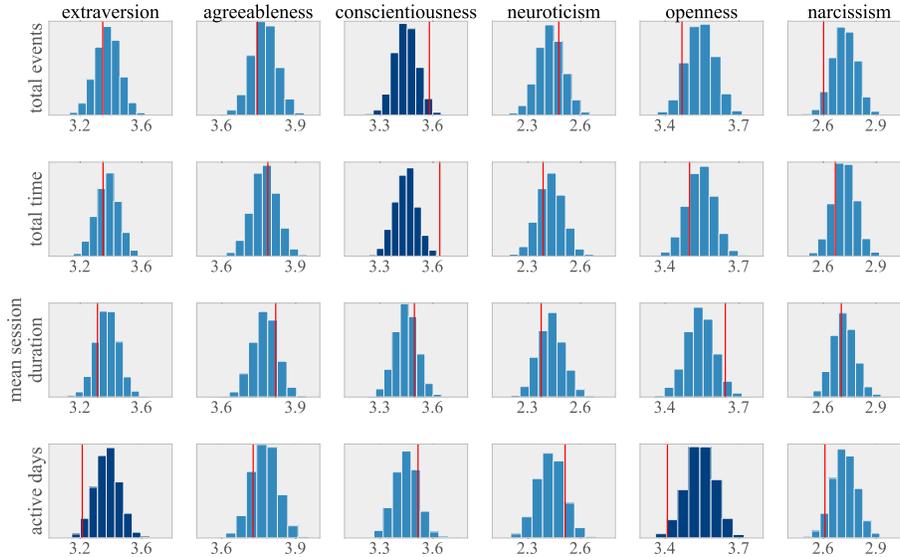}
\caption{
Distributions of the bootstrapped means of the subsamples. 
The red line is the measured mean of the top 10\% quantile for each trait.
The cases where the mean of top 10\% are more extreme than 95\% of the bootstrapped samples are highlighted in dark blue.
}
\label{distributions}
\end{figure}

Fig.~\ref{distributions} depicts the bootstrapping distributions of each personality trait and usage feature. The red line indicates the measured mean of the top 10\%. 
We note that, as before, conscientiousness with total time have the most visible difference from the bootstrapped means, but also some for other feature-trait pairs (highlighted in dark blue) there are quite large differences.

\section*{Discussion}
A key area of interest in this work is to understand the reasons for usage and non-usage of the SensibleJournal 2014 app as a Personal Informatics self-tracking system. 
Our population consists of well-educated, tech-savvy young adults, a fact that may introduce some biases. 
However, we suggest that this population would be possible adopters of self-tracking tools.

The total usage varies greatly from participant to participant. 
About 30\% of the participants did not use the app at all and only a small fraction used it every few days. 
Since significant amount of time and resources are spent in developing such Personal Informatics systems, this inability to engage the potential users can be a concern for any such system, either within a research project in academia, or a product in commercial settings. 
A possible factor for the limited uptake of the app is the fact that the participants might receive similar knowledge from other commercial and more visually polished apps or services, thus resulting in rapid loss of interest.
% NOTE: Removed: If we look at the top 10\% of the participants in terms of usage, we see that, on average, they were launching the app more frequently and with shorter break periods than the rest.

The interest was higher in the initial phase, and then declined as time progressed. 
The decrease over time may have multiple explanations, such as the repetitive nature of the feedback. 
As many respondents reported, the app tends to report similar information over time, that is time spent usually at the same few places, like home and work. 
Therefore, users may not be able to learn anything new and would eventually abandon the app. 
The repetitive feedback information is due to the inherent nature of human mobility, which tends to be habitual and predictable \cite{song2010limits}. 
Habitual living suggests that once initial insights have been obtained, limited new significant knowledge can be gained from the data itself. 
This problem may potentially affect many Personal Informatics systems measuring periodic behavior such as fitness activity, heart rates and sleep patterns. 
Any app reporting about the status of the user will soon produce repetitive feedback and may risk to become uninteresting for the user.

One possible solution is to generate feedback only when new or deviating information is available, such as something that has not happened before, something that is different or something requiring user attention. 
A goal-setting feature could also stimulate the interest in self-reflection and facilitate the process of behavior change.

One hypothesis was that personality traits are a factor in the adoption of Quantified Self tools. 
We find, however, that narcissism makes no significant difference in respect to adoption, in contrast to conscientiousness, which is the only trait making a statistically significant difference. 
The present data is insufficient for a full understanding of the casual relationship, but we hypothesize that the organization and self-discipline characteristics of conscientiousness could be an important driver for the usage of such self-tracking apps.

\section*{Conclusions}
We have presented results from an empirical study on the usage of a mobility self-tracking smartphone app, among $N$=796 participants in an ongoing mobile sensing study. 
The app was offered to all participants, but whether they would use it or not was left optional and not enforced in any way. 

A relatively low uptake of the app was observed, as 30\% of the participants never used the app. 
Only 16 participants (2\%) ended up using the app on a regular basis and among that group, a decline in usage over time was observed. 
A questionnaire on the participants' interests and attitudes towards the self-tracking app provided indications that the recurring data patterns lead to a drop in interest, once initial insights on personal mobility had been obtained. 
The fact that the app does not provide radically different information over time, or suggestions based on the data obtained, is a possible reason for the participants’ gradual loss of interest in the app, as well as in their personal mobility data.

Personality traits of the $N$=796 participants were corresponding to those of the general population. 
In order to understand potential differences between those who adopted the self-tracking behavior by using the app and those who decided not to use it (or only use it short-term), we compared their personality traits. 
Through this comparison, we found a relation between self-tracking and conscientiousness, an observation that is in contrast with the view in popular media, which suggest a tendency towards narcissism among people that adopt self-tracking behavior.

\section*{Acknowledgments}
AC's work is funded in part by the \emph{High Resolution Networks} project (The Villum Foundation), as well as \emph{Social Fabric} (University of Copenhagen).
GC's work was performed while under PhD fellowship at Athena Research and Innovation Center, Athens, Greece.

\bibliographystyle{abbrv}
\bibliography{bibliography}

\end{document}